\documentclass[aps, prl, twocolumn,preprintnumbers,amsmath,amssymb,longbibliography]{revtex4-1}

\usepackage{graphicx}% Include figure files
\usepackage{dcolumn}% Align table columns on decimal point
\usepackage{bm}% bold math
\usepackage{hyperref}

\begin{document}

\title{A two-neutron halo  is unveiled in $^{29}$F}

\author {S. Bagchi$^{1,2,3}$, R. Kanungo$^*$$^{1,4}$, Y. K. Tanaka$^{1,2,3}$, H. Geissel$^{2,3}$, P. Doornenbal$^{5}$, W. Horiuchi$^6$, G. Hagen$^{7,8}$, T. Suzuki$^9$, N. Tsunoda$^{10}$, D. S. Ahn$^{5}$,  H. Baba$^{5}$,  K. Behr$^2$, F. Browne$^{5}$, S. Chen$^{5}$,  M.L. Cort\'es$^{5}$,  A. Estrad\'e$^{11}$, N. Fukuda$^{5}$, M. Holl$^{1,4}$, K. Itahashi$^{5}$,  N. Iwasa$^{12}$,  G.R. Jansen$^{7,13}$, W. G. Jiang$^{8,7}$, S. Kaur$^{1,14}$, A.O. Macchiavelli$^{15}$, S. Y. Matsumoto$^{16}$,  S. Momiyama$^{17}$,  I. Murray$^{5,18}$,   T. Nakamura$^{19}$,  S. J. Novario$^{8,7}$, H.J. Ong$^{20}$, T. Otsuka$^{5,17}$, T. Papenbrock$^{8,7}$, S. Paschalis,$^{21}$  A. Prochazka$^2$, C. Scheidenberger$^{2,3}$, P. Schrock$^{22}$, Y. Shimizu$^{5}$,  D. Steppenbeck$^{5,22}$, H. Sakurai$^{5,17}$, D. Suzuki$^{5}$,  H. Suzuki$^{5}$, M. Takechi$^{23}$, H. Takeda$^5$,  S. Takeuchi$^{19}$, R. Taniuchi$^{17,21}$, K. Wimmer$^{17}$,  K. Yoshida$^{5}$ }

\affiliation{$^1$Astronomy and Physics Department, Saint Mary's University, Halifax, NS B3H 3C3, Canada}
\affiliation{$^2$GSI Helmholtzzentrum f\"ur Schwerionenforschung GmbH, D-64291 Darmstadt, Germany}
\affiliation{$^3$Justus-Liebig University,  35392 Giessen, Germany}
\affiliation{$^4$TRIUMF, Vancouver, BC V6T 2A3, Canada}
\affiliation{$^{5}$RIKEN Nishina Center, Wako, Saitama 351-0198, Japan}
\affiliation{$^6$Department of Physics, Hokkaido University, Sapporo 060-0810, Japan}
\affiliation{$^7$Physics Division, Oak Ridge National Laboratory, Oak Ridge, TN 37831, USA}
\affiliation{$^8$Department of Physics and Astronomy, University of Tennessee, Knoxville, TN 37996, USA}
\affiliation{$^9$Department of Physics, Nihon University, Setagaya-ku, Tokyo 156-8550, Japan}
\affiliation{$^{10}$Center for Nuclear Study, University of Tokyo, Bunkyo-ku, Tokyo 113-0033, Japan}
\affiliation{$^{11}$Department of Physics, Central Michigan University, Mount Pleasant, MI 48859, USA}
\affiliation{$^{12}$Department of Physics, Tohoku University, Miyagi, 980-8577, Japan}
\affiliation{$^{13}$National Center for Computational Sciences, Oak Ridge National Laboratory, Oak Ridge, Tennessee 37831, USA}
\affiliation{$^{14}$Department of Physics and Atmospheric Science, Dalhousie University, Halifax, NS B3H 4R2, Canada}
\affiliation{$^{15}$Nuclear Science Division, Lawrence Berkeley National Laboratory, Berkeley, California 94720, USA}
\affiliation{$^{16}$Department of Physics, Kyoto University, Kyoto 606-8502, Japan}
\affiliation{$^{17}$Department of Physics, University of Tokyo, Bunkyo-ku, Tokyo 113-0033, Japan}
\affiliation{$^{18}$Institut de Physique Nucleaire, IN2P3, CNRS, Universit\'e Paris-Sud, Universit\'e Paris-Saclay, 91406 Orsay Cedex, France}
\affiliation{$^{19}$Department of Physics, Tokyo Institute of Technology, 2-12-1 O-Okayama, Meguro, Tokyo 152-8551, Japan}
\affiliation{$^{20}$RCNP, Osaka University, Mihogaoka, Ibaraki, Osaka 567 0047, Japan}
\affiliation{$^{21}$Department of Physics, University of York, Heslington, York YO10 5DD, UK}
\affiliation{$^{22}$Center for Nuclear Study, University of Tokyo, RIKEN Campus, Wako, Saitama 351-0198, Japan}
\affiliation{$^{23}$Graduate School of Science and Technology, Niigata University, Niigata 950-2102, Japan}

\begin{abstract}
We report the measurement of reaction cross sections ($\sigma_R^{\rm
  ex}$) of $^{27,29}$F with a carbon target at RIKEN. The unexpectedly
large $\sigma_R^{\rm ex}$ and derived matter radius identify $^{29}$F
as the heaviest two-neutron Borromean halo to date. The halo is
attributed to neutrons occupying the $2p_{\rm 3/2}$ orbital, thereby
vanishing the shell closure associated with the neutron number $N$ =
20. The results are explained by state-of-the-art shell model
calculations. Coupled-cluster computations based on effective field
theories of the strong nuclear force describe the matter radius of
$^{27}$F but are challenged for $^{29}$F.
\end{abstract}

\maketitle

In atomic nuclei the strong force binds protons and neutrons into
complex systems. Long-lived isotopes and $\beta$-stable nuclei exhibit
a well-known shell structure~\cite{mayer1950,haxel1949}.  However, in
some nuclei with a large neutron excess an unusual type of structure
emerges. In neutron-halo nuclei a large nuclear surface is formed that
is almost entirely composed of
neutrons~\cite{tanihata1985,hansen1987}. Particularly interesting are
so-called Borromean two-neutron halos \cite{vaagen2000}.  These
intriguing quantum systems consist of a bound state between a core
nucleus and two neutrons, where any of the two-body sub-systems are
unbound. Examples known so far are $^6$He, $^{11}$Li, $^{14}$Be,
$^{17}$B, and $^{22}$C.  A neutron-halo nucleus exhibits an enhanced
root-mean-square matter radius ($R_m^{\rm ex}$) that can be extracted
from the (unusually large) reaction cross section $\sigma_R^{\rm ex}$,
which deviates from the known trend $R_m^{\rm ex}\propto A^{1/3}$ with
mass number $A$. Some general conditions for halos are summarized in
Ref.~\cite{jensen2000}. These exotic nuclei are intricately related to
changes in the nuclear shell structure. In $^{11}$Li, for example, the
$N$ = 8 shell gap vanishes with the intruder 2$s_{\rm 1/2}$ orbital
(35 - 55\%) that forms a Borromean halo in the last bound
isotone~\cite{simon1999,tanihata2008}.

Do all traditional neutron shell closures vanish into Borromean
two-neutron halos? We address this question here for $N$ = 20 by
reporting the discovery of the heaviest Borromean halo to date, and
the first of its kind in the proton $sd$-shell.  The measured total
reaction cross section $\sigma_R^{\rm ex}$ of the $N$ = 20 nucleus
$^{29}$F is much larger than that of $^{27}$F. This observation
implies a two-neutron halo structure in $^{29}$F, and the
corresponding melting of the traditional $N$ = 20 shell gap is due to
the intrusion of the 2$p_{\rm 3/2}$ orbital from a higher
shell. Therefore, the two weakly bound neutrons experience only a
small centrifugal barrier and have extended wavefunction to form the
halo.

The weakening of the $N$ = 20 shell gap was first hinted at from
systematics of the two-neutron separation energies ($S_{2n}$) of
sodium isotopes \cite{thibault1975} and subsequently observed through
the low excitation energy \cite{detraz1979} and enhancement of reduced
electric quadrupole transition probability \cite{motobayashi1995} of
$^{32}$Mg. Since then a large number of investigations in neon to
aluminum isotopes found intruder $pf$-shell components in level
schemes \cite{tripathi2005,doornenbal2009}, orbital configurations
\cite{takechi2012,nakamura2014,kobayashi2016,liu2017}, and magnetic
moment \cite{neyens2005}.

Monte Carlo Shell Model calculations~\cite{utsuno1999} align well with
these findings. It suggests that the monopole tensor interaction
contributes to the shell quenching~\cite{otsuka2005,otsuka2020}. The
high atomic number ($Z$) boundary of the quenched shell is drawn at
the aluminum isotopes.  The low-$Z$ shore of this quenched shell
remains undetermined. The observed lowest resonance state of $^{28}$F
can be explained by the USDB shell model interaction without
appreciable need for any intruder orbitals from the $pf$
shell~\cite{christian2012} thereby concluding it to follow normal
shell ordering. Large-scale shell model calculations, however, predict
the Borromean nucleus $^{29}$F to be at the boundary of normal to
quenched shells \cite{caurier2014}.

The boundary of bound nuclear landscape, the drip-lines, are defined
by the last bound isotopes or isotones \cite{hansen2003}. We have few
data on nuclei close to the neutron-drip line of the $N$ = 20
isotones.  In $^{29}$F, the two-neutron separaton energy $S_{2n} =
1.4(6)$~MeV is only known with a low precision
\cite{gaudefroy2012}. The excited states of $^{27,29}$F are observed
\cite{doornenbal2017} at 915(12)~keV and 1080(18)~keV,
respectively. The state in $^{29}$F is slightly higher than
shell-model prediction using the SDPF-M interaction~\cite{utsuno1999}
that includes the $pf$ shell.  A particle-rotor picture
\cite{macchiavelli2017} also explains the $^{29}$F spectrum, using a
deformed $^{28}$O core coupled to a proton in the 1$d_{5/2}$ Nilsson
multiplet.  Regarding neutron halos, our knowledge is similarly
limited.  Carbon is the last known element to exhibit a Borromean
two-neutron halo, and we do not know about any neutron halos in
fluorine.

\begin{figure}
  \includegraphics[width=8.9cm, height=7cm]{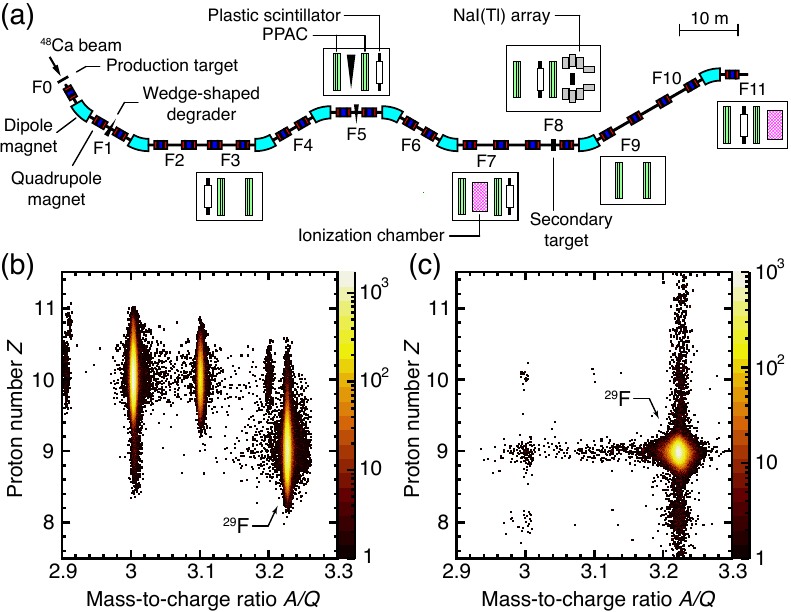}
  \caption{\label{fig:1} (a) Schematic view of the experimental
    setup. The nuclei $^{27,29}$F are transported from the focal plane
    F0 to F8, where the reaction target is located. Unreacted $^{29}$F
    is identified using the ZDS from F8 to F11. Particle
    identification (b) before the carbon reaction target at F8 and (c)
    after the target at F11 with $^{29}$F events selected before the
    target. }
\end{figure}

In this work, we report on the first measurement of the interaction
cross sections ($\sigma_{I}^{\rm ex}$) and determination of point
matter radius of $^{27,29}$F. The experiment was performed at the
Radioactive Isotope Beam Factory operated by the RIKEN Nishina Center
and the Center for Nuclear Study (CNS), University of Tokyo, Japan
using the BigRIPS fragment separator and ZeroDegree spectrometer
(ZDS)~\cite{kubo2012}. The experimental setup is shown in
Fig.~\ref{fig:1}a. The $^{27,29}$F isotopes were produced from
fragmentation of a $^{48}$Ca beam with an average intensity of 570~pnA
and an energy of 345$A$~MeV interacting with a 10~mm thick rotating Be
target. The isotopes of interest were separated from the various
contaminant fragments using the BigRIPS fragment separator and
identified [Fig.~\ref{fig:1}(b)] using the technique of in-flight
energy deposit ({\it {$\Delta$E}}), time of flight (TOF) and magnetic
rigidity ({\it {B$\rho$}}).  Achromatic wedge shaped aluminium
degraders of thicknesses 15~mm and 5~mm were used at the dispersive
foci F1 and F5 [black inverted triangle in Fig.~\ref{fig:1}(a)],
respectively, to spatially separate the beam contaminants. The $B
\rho$ was determined from a position measurement with parallel plate
avalanche counters (PPAC)~\cite{kumagai2013} placed at the F3, F5 and
F7 focal planes [green boxes in Fig.~\ref{fig:1}(a)]. An ionization
chamber placed at F7 [pink box in Fig.~\ref{fig:1}(a)] provided the
{\it {$\Delta$E}} information. Plastic scintillator detectors of 3 mm
thickness located at the F3 and F7 focal planes (white boxes in
Fig.1a) provided the TOF information.  A $2.01 \pm 0.01$~g/cm$^2$
thick carbon reaction target was placed at F8 and was surrounded by
the DALI2 NaI(Tl) array \cite{takeuchi2014} for detecting gamma rays
from the reactions. The average beam rates onto the F8 target were
314~pps and 78~pps, whereas the beam energies before the F8 target
were 250$A$~MeV and 255$A$~MeV for $^{27}$F and $^{29}$F,
respectively. In the event selection of fluorine isotopes, the
relative contribution from Ne isotopes was $\leq 2\times 10^{-9}$.

The $\sigma_{I}^{\rm ex}$ of the $^{A}$F nuclei were measured via the
transmission technique where the number of the incident nuclei
($N_{\rm in}$) is obtained from an event-by-event counting at F7 and
F8. After interaction with a carbon reaction target at F8, the
unreacted $^A$F ($N_{\rm out}$) were counted at the F11 focal
plane. The $\sigma_{I}$ was then obtained from the relation
$\sigma_{I}^{\rm ex}=t^{-1}\ln(T_{{\rm t-out}}/T_{{\rm t-in}})$, where
$T_{{\rm t-in}}$ and $T_{{\rm t-out}}$ are the ratios of $N_{\rm
  out}/N_{\rm in}$ with and without the reaction target, respectively
and $t$ is the areal thickness of the target. Empty-target
measurements were needed in order to take into account the losses due
to interactions with residual materials in the beam-line and detection
efficiencies. Constant transmission throughout the ZDS was obtained by
restricting the phase space in X, Y and momentum directions before the
reaction target at F8.

The unreacted $^{27,29}$F residues were analyzed using the ZDS. The
{\it {$\Delta$E}} of these ions was measured using a Multi-Sampling
Ionization Chamber (MUSIC)~\cite{stolz2002} detector [pink box in
  Fig.~\ref{fig:1}(a)] placed at the final achromatic focal plane F11
of the ZDS. The TOF was measured between two plastic scintillators
having thicknesses 3~mm and 1~mm placed at the achromatic focal planes
F8 and F11, respectively. The B$\rho$ was determined from the PPACs
placed at the dispersive focal plane F9 and final focus
F11. Figure~\ref{fig:1}(c) shows the particle identification obtained
in the ZDS for events selected as $^{29}$F before the reaction target
at F8. The resolution of $Z$ is obtained to be 0.2 (FWHM) and that of
$A/q$ for the F isotopes is 0.013 (FWHM).

\begin{figure}
\includegraphics[width=7 cm, height=6.5 cm]{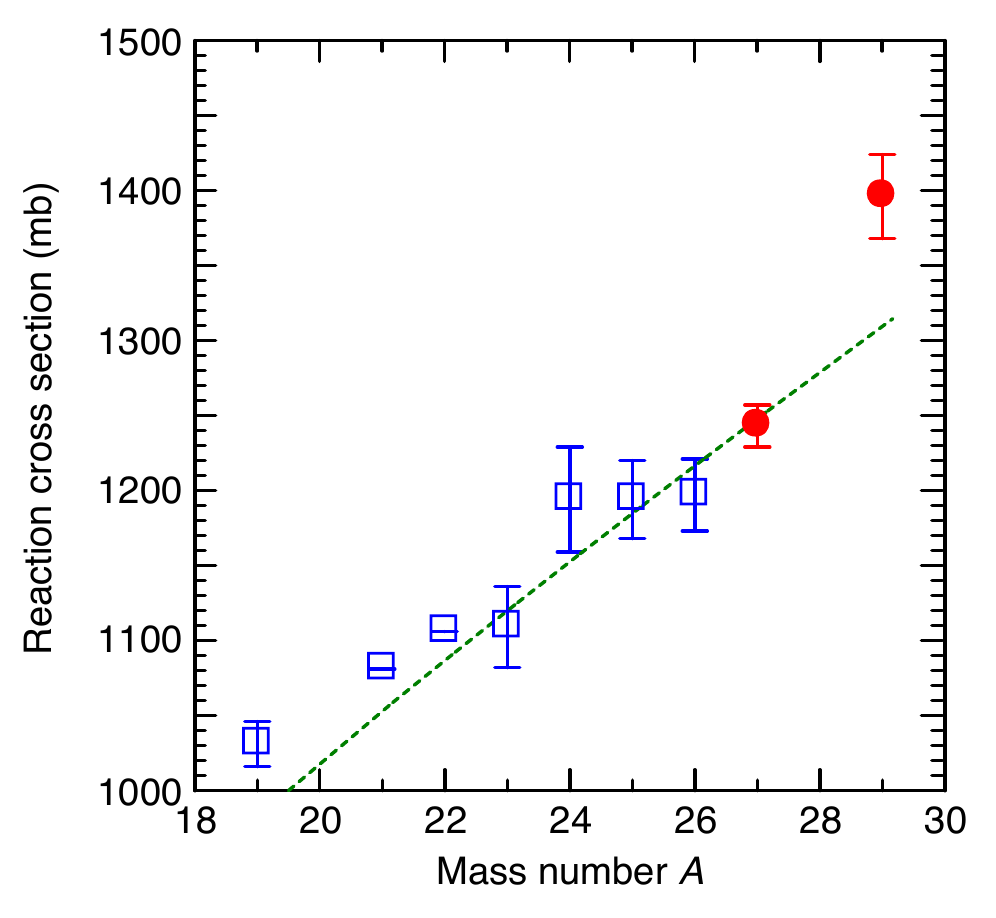}
\caption{\label{fig:2} Measured reaction cross sections of fluorine
  isotopes with a carbon target at $E/A \approx$ 240 MeV. The red
  filled circles are data of the present work. The open blue squares
  are from Ref.~\cite{homma2017}. The data show statistical and
  systematic uncertainties.  The dashed line shows the trend of
  $A^{2/3}$ relative evolution normalized for best fit to
  $^{19-27}$F.}
\end{figure}

The reaction cross section $\sigma_{R}^{\rm ex}$ is the sum of
$\sigma_{I}^{\rm ex}$ and the inelastic scattering cross section
($\sigma_{\rm{inel,bs}}$) to bound excited states. No gamma rays from
inelastic scattering were observed.  The efficiency of 1 MeV
$\gamma$-ray detection was $\sim$20\%. The inelastic scattering
$\gamma$-ray spectrum in Ref.\cite{togano2016} for $^{20}$C yields a
cross section of $\sim$ 3 mb. Therefore, non-observation of a
$\gamma$-ray peak places an upper limit of $\sigma_{\rm{inel,bs}}$ to
less than 1~mb for $^{27,29}$F.  Hence, $\sigma_{R}^{\rm
  ex}\approx\sigma_{I}^{\rm ex}$. The $\sigma_R^{\rm ex}$ for
$^{27,29}$F, 1243(14) mb and 1396(28) mb, respectively (red filled
circles) and those for $^{19-26}$F from Ref.~\cite{homma2017} (open
blue squares), presented in Fig.~\ref{fig:2}, show a steep increase of
about 12(2)\% for $^{29}$F revealing the presence of a two-neutron
halo. This increase in $\sigma_R^{\rm ex}$ is similar to that found
for $^{22}$C \cite{togano2016}.

The $\sigma_R$ are calculated from the Glauber model with the
nucleon-target profile function and a harmonic oscillator density for
the $^{12}$C target (see Supplemental Material for further
details). For $^{27,29}$F we consider harmonic oscillator densities
with several oscillator width parameters that yield different
point-matter radii for these nuclei.  Using each of these densities we
evaluate the $\sigma_R$ with the Glauber model. The calculated
$\sigma_R$ are compared to the measured $\sigma_R^{\rm ex}$ to extract
the experimental $R_m^{\rm ex}$ of $3.15 \pm 0.04$~fm and $3.50 \pm
0.07$~fm for $^{27}$F and $^{29}$F, respectively. The results obtained
are also consistent with a two-parameter Fermi density function. The
large increase of $R_m^{\rm ex}$ by about 11(3)\% for $^{29}$F
compared to $^{27}$F is consistent with a two-neutron halo formation
in the $N$ = 20 isotone at the drip line and is well above the 2.4\%
increase expected from the $A^{1/3}$ rule. A large root-mean-square
halo radius of 6.6~fm for $^{29}$F is derived considering the proton
radii in $^{27}$F and $^{29}$F to be similar. The difference between
the $R_m^{\rm ex}$ of $^{29}$F and its core $^{27}$F is 0.35$\pm$0.08
fm which is similar to the two-neutron halo nuclei $^{14}$Be, $^{17}$B
\cite{ozawa2001} and $^{22}$C \cite{togano2016}.

To assess the neutron orbitals associated with the halo, we perform
Glauber calculation with a density of $^{29}$F as $^{27}$F+n+n. The
large increase of the matter radii from $^{27}$F to $^{29}$F indicates
a strong component of the intruder 2$p_{3/2}$ orbital. Its centrifugal
barrier being a factor of three lower than the 1$d_{3/2}$ orbital
facilitates an extended wavefunction. The large extension becomes
possible due to the small S$_{2n}$ in $^{29}$F \cite{gaudefroy2012}
approaching the effective threshold as shown for higher angular
momentum orbital in Ref.\cite{hoffman2016}.  To obtain the $^{29}$F
density, we assume mixing of the (1$d_{3/2})^2$ and (2$p_{3/2})^2$
configurations with their wavefunctions generated from the Woods-Saxon
potential using a single-neutron energy of $S_{2n}/2= 0.7(3)$~MeV
\cite{gaudefroy2012} (see Supplemental Materials for more
details). Figure~\ref{fig:3} shows the result of the mixing according
to $\sigma_R$ = $\alpha \times \sigma_R({\rm{2}}p_{3/2})+(1-\alpha)
\times \sigma_R({\rm{1}}d_{3/2})$ with $\alpha$ being the occupation
probability normalized to unity. For $^{29}$F, the consistency between
$\sigma_R^{\rm ex}$ and the $\sigma_R$ calculated with the Glauber
model requires $\alpha$ = 0.54 - 1.0 for $S_{2n}/2$ = 0.7 MeV,
indicating that the halo is driven by the lowering of the 2$p_{3/2}$
orbital and the $N$ = 20 and 28 shell closures vanishing.  The
uncertainty in $S_{2n}$ gives a lower limit of $\alpha$= 0.36
(Fig. 3).  One can also describe $^{27}$F as a $^{26}$F+n
configuration (where the $^{26}$F core radius $R_m$ is taken to
reproduce its $\sigma_R^{\rm ex}$). In this approach, the neutron
occupation in the 1$d_{3/2}$ orbital alone is able to explain
$\sigma_R^{\rm ex}$ of $^{27}$F, suggesting a very small contribution
of the intruder $pf$ orbitals.

\begin{figure}
  \includegraphics[width=7 cm, height=6.5 cm]{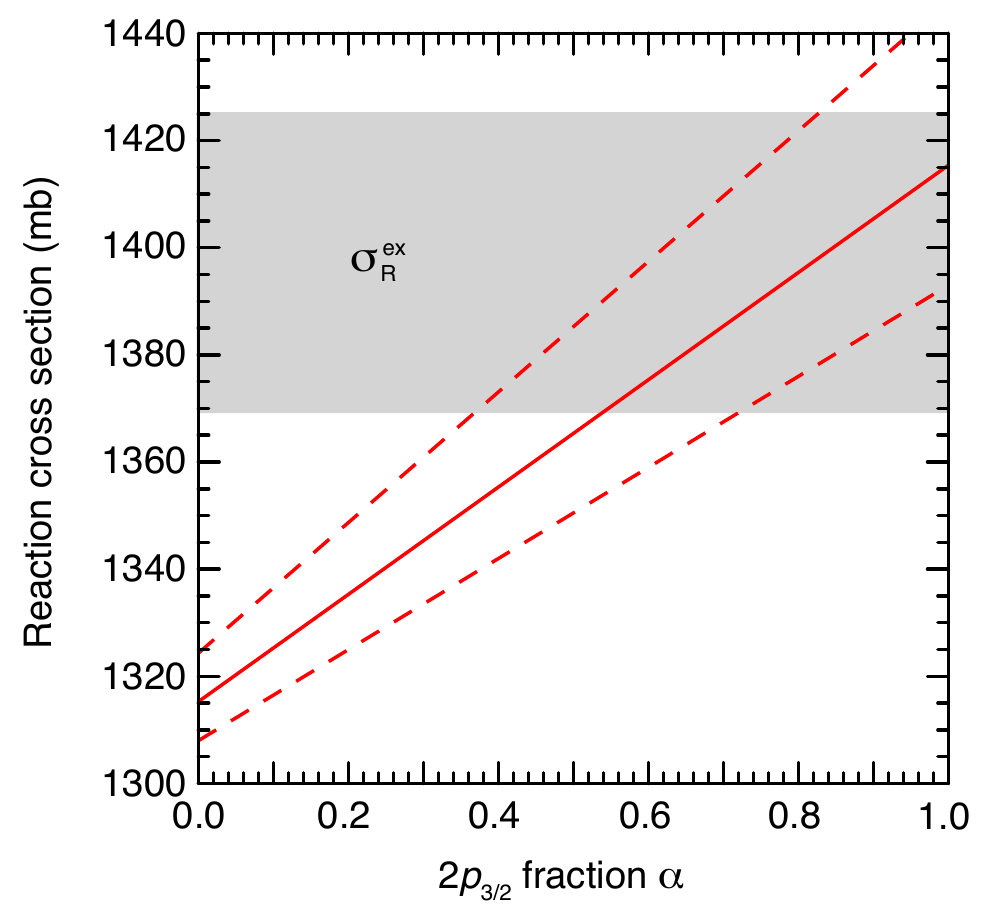}
  \caption{\label{fig:3} The red lines show Glauber calculation of
    $\sigma_R$ of $^{29}$F+C with $^{27}$F+n+n densities at E/A = 246
    MeV (mid-target energy) for different fractions ($\alpha$), ({\it
      {see text}} for definition) of neutrons in the 2$p_{3/2}$
    orbital. The solid (dashed) lines are for $S_{2n}$/2= 0.7(3)
    MeV. The horizontal shaded band corresponds to the measured
    $\sigma_R^{\rm ex}$. }
\end{figure}

\begin{figure}
\includegraphics[width=7 cm, height=12 cm]{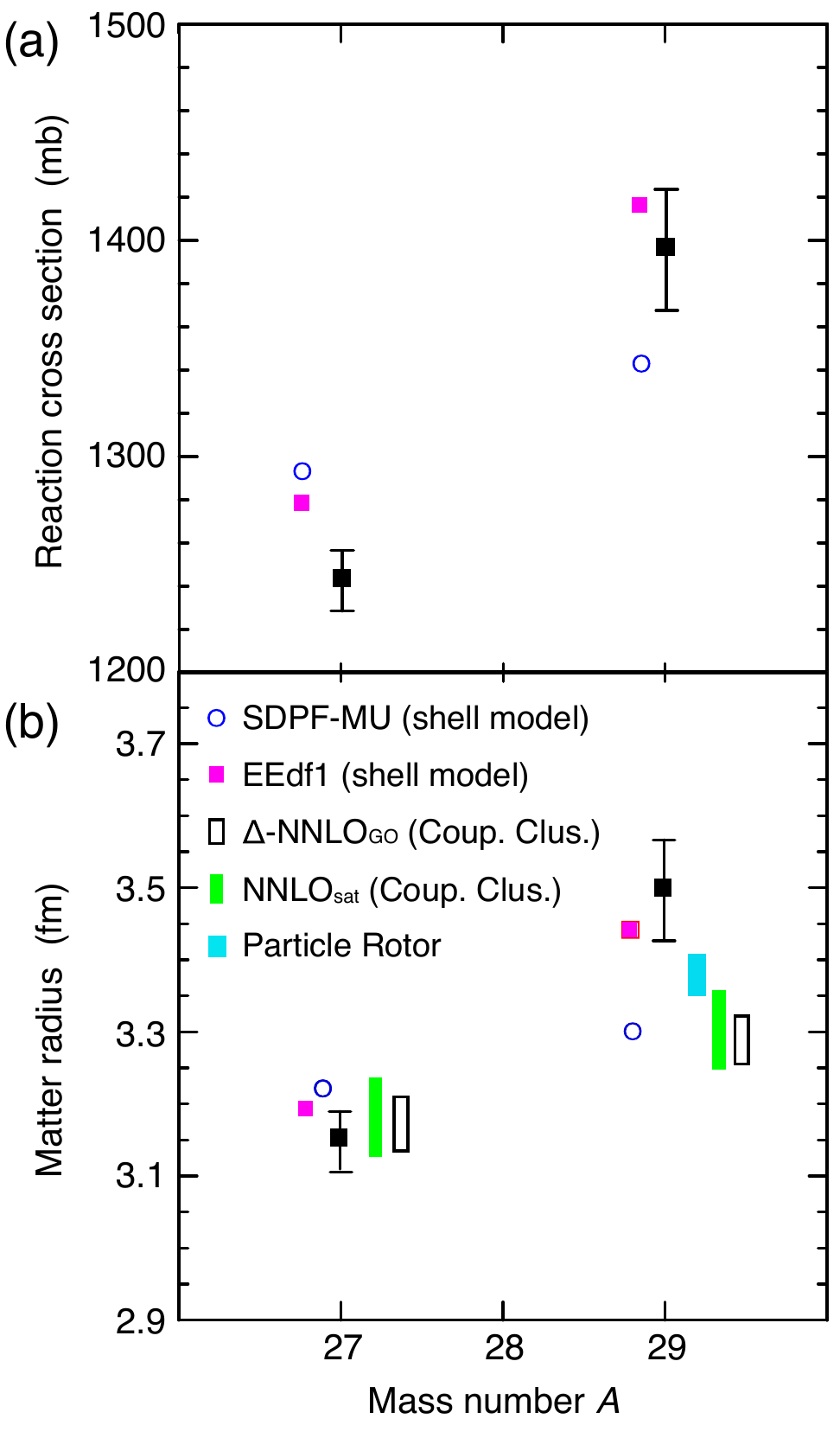}
\caption{\label{fig:4} Comparison between the experimental values for
  $\sigma_R^{\rm ex}$ and $R_m^{\rm ex}$ and various theoretical
  results.  The data include statistical and systematic uncertainties.
  (a) Comparison of measured $\sigma_R^{\rm ex}$ (black filled
  squares) with Glauber calculations using density distributions from
  the shell model based on the SDPF-MU (open blue circles) and the
  EEdf1 (magenta square) interaction. (b) Comparison of derived
  $R_m^{\rm ex}$ for $^{27,29}$F (black filled squares) with
  predictions from shell-model calculations using the SDPF-MU
  interaction (open blue circles) and the EEdf1 interaction (magenta
  squares). Coupled-cluster results based on the chiral
  NNLO$_{\rm{sat}}$ ($\Delta$-NNLO$_{\rm GO}$) interaction are shown
  as a green band (open black-white band). The cyan band is result
  from a particle-rotor model. }
\end{figure}

In order to gain further insight into the shell structure driving the
halo formation, the matter radii are evaluated by using occupation
numbers obtained from shell-model calculations in the $sd$-$pf$ shell.
One calculation is performed with the SDPF-MU
Hamiltonian~\cite{utsuno2012}.  For $^{27,29}$F, radial wave functions
are calculated in a Woods-Saxon potential (see Supplemental Material
for further details). The $\sigma_R$ using these densities for
$^{27,29}$F are shown by open blue circles in Fig.~\ref{fig:4}(a). The
resultant matter radii are 3.22~fm and 3.30~fm (open blue circles in
Fig.~\ref{fig:4}(b)) for $^{27}$F and $^{29}$F, respectively. The
corresponding neutron occupation numbers of the 1$d_{3/2}$,
1$f_{7/2}$, and 2$p_{3/2}$ orbitals are predicted as 2.68, 0.90, and
0.56 in $^{29}$F and 1.67, 0.48, and 0.24 in $^{27}$F. The
underprediction of $R_m$ and $\sigma_{\rm R}$ for $^{29}$F can be
traced back to unbound $pf$ orbitals. These appear with a small
component in the ground-state configuration while the 1$d_{3/2}$
orbital is bound and has a larger component. The predicted first
excited states in $^{27,29}$F are at 1.48 MeV and 1.51 MeV,
respectively, slightly higher than the data in
Ref.\cite{doornenbal2017}.

Matter radii and $\sigma_R$ are also evaluated with a microscopic
interaction called EEdf1 \cite{Dom地guez2018} which has been
derived \cite{tsunoda2017} by the extended Kuo-Krenciglowa (EKK)
method
\cite{krenciglowa1974,takayanagi2011,takayanagi2011a,tsunoda2014} from
a chiral N$^3$LO interaction~\cite{entem2003} and Fujita-Miyazawa
three-body force~\cite{fujita1957} (magenta squares in
Fig.~\ref{fig:4}). The $sd$- and $pf$-shells are more strongly mixed
than by the SDPF-MU interaction, with neutron occupation numbers of
the 1$d_{3/2}$, 1$f_{7/2}$ and 2$p_{3/2}$ orbitals in $^{29}$F
($^{27}$F) being 0.84, 2.19, and 1.26 (0.80, 1.08, and 0.67),
respectively. The substantial contribution of the bound 2$p_{3/2}$
orbital leads to the observed halo formation.  The computed matter
radius of 3.44~fm for $^{29}$F agrees with the data, while that of
$^{27}$F is 3.19~fm [magneta squares in Fig.~\ref{fig:4}(b)]. We note
that the 1$d_{3/2}$ orbital is unbound with the EEdf1 interaction. It
predicts the first and second excited states in $^{27}$F at 0.14 MeV
and 1.42 MeV, respectively.  Those in $^{29}$F are predicted at 0.09
MeV and 1.08 MeV, respectively, the latter being in agreement with the
observed $\gamma$-ray transition \cite{doornenbal2017}. This suggests
the first excited state could be below the detection threshold. The
low excitation energies in $^{29}$F align with the quenching of the
$N$ = 20 shell closure.

We also performed {\it ab-initio} coupled-cluster
calculations~\cite{coester1960,kuemmel1978,bishop1991,mihaila2000b,dean2004,bartlett2007,hagen2014}
for the binding energies and matter radii of $^{27,29}$F. These
computations are based on a deformed reference state. We used two
different interactions from chiral effective field
theory~\cite{vankolck1999,epelbaum2009,machleidt2011} that consist of
nucleon-nucleon and three-nucleon forces, namely NNLO$_{\rm
  sat}$~\cite{ekstrom2015} and $\Delta$-NNLO$_{\rm
  GO}$(450)~\cite{jiang2019}. Both interactions are constrained by
nuclear saturation properties (see Supplemental Material for further
details). The coupled-cluster results for the matter radius of
$^{27}$F agree with the data, while those for $^{29}$F are smaller
than the data. Error ranges reflect uncertainties with respect to
model-space sizes and extrapolation of radii. In our deformed
reference state, the neutrons closest to the Fermi surface occupy
positive parity states, dominantly associated with the $1d_{3/2}$
orbital. These states were self-consistently selected by the
Hartree-Fock method.  A halo in $^{29}$F would require neutrons to
occupy the $2p_{3/2}$ orbital.  Thus, the coupled-cluster computations
lead to smaller radii pointing to short comings in the employed
interactions.

The matter radius is also estimated for $^{29}$F using the
particle-rotor model \cite{macchiavelli2017}, assuming a prolate
ellipsoidal shape.  This approach hints to a possible effective
deformed $^{28}$O core (with deformation $\varepsilon_2 \approx
0.16^{+0.15}_{-0.2}$) \cite{macchiavelli2017}, supportive of a
breakdown of the $N$ = 20 shell. The resulting radius is slightly
lower than the data (cyan bar in Fig.~\ref{fig:4}).

While this article was under review, the matter radii of $^{27-31}$F
were predicted {in a Gamow Shell Model framework
  \cite{Michel2020}. The prediction for $^{27}$F is slightly higher
  and that of $^{29}$F is slightly lower than the data presented
  here. Future experiments will aim to assess the halo predicted for
  $^{31}$F in Ref.\cite{Michel2020} and a pairing antihalo effect
  predicted in Ref.\cite{Masui2020}.

The present work shows that a small neutron separation energy ($\sim$
1 MeV), and tensor force effects lead to a $p$-wave halo in $^{29}$F,
one proton above conventional doubly closed shell $Z$ = 8 and $N$ =
20.  This is analogous to $s$-wave halo in $^{11}$Li, one proton above
$Z$ = 2 and $N$ = 8.  Both $^{29}$F and $^{11}$Li are at the neutron
drip-line with the respective conventional doubly-magic cores,
$^{28}$O and $^{10}$He, being unbound. The extended wavefunctions of
such weakly bound $s$ or $p$ orbitals in the ground states of nuclei
around the $N$ = 50, 82 and 126 shells will lead to greater
probability of neutron capture \cite{mathews1983} thereby impacting
the flow of the rapid neutron capture process.  One-neutron halos and
quenching of the $N$ = 50 shell gap are predicted in Cr and Fe
isotopes \cite{hamamoto2017} and two-neutron halo in Ca isotopes
\cite{hagen2013}. A recent study of $^{207}$Hg beyond $N$ = 126 shows
the normal shell ordering to persist \cite{Tang2020}. Calculations
with a Woods-Saxon potential however predict a shell gap quenching due
to weak binding in more neutron-rich $N$ = 126 isotones. This follows
the trend in light nuclei discussed in
Refs. \cite{ozawa2000,hoffman2014}.

In conclusion, we identified a new two-neutron Borromean halo -- the
first of this kind in the proton $sd$-shell -- in the $N$ = 20
drip-line nucleus $^{29}$F. This observation was from the large
difference in the reaction cross sections $\sigma_R^{\rm ex}$ measured
for $^{27,29}$F. Assuming similar proton distributions in $^{27}$F and
$^{29}$F yields a large root-mean-square halo radius of 6.6~fm for
$^{29}$F. The emergence of the halo leads to vanishing of the $N$ = 20
shell closure with contribution of the 2$p_{3/2}$ orbital. This
weakens the $N$ = 28 shell gap as well. The disappearance of the
conventional shell gap and emergence of the halo challenges {\it ab
  initio} computations and will trigger further experiments
characterizing this halo.

The authors gratefully thank the RIKEN Rare Isotope Beam Factory for
delivering the $^{48}$Ca beam with unprecedented high intensity. The
support from NSERC Canada is gratefully acknowledged. This work was
also supported by JSPS KAKENHI Grant Nos. JP16H02179, JP18H05404, UK
STFC under contract number ST/P003885/1, the U.S. Department of
Energy, Office of Science, Office of Nuclear Physics under Contract
No. DE-AC02-05CH11231 (LBNL). The work was supported in part by the
Office of Nuclear Physics, U.S. Department of Energy, under grants
DE-SC0018223 (SciDAC-4 NUCLEI collaboration), DE-FG02-96ER40963, and
by the Field Work Proposal ERKBP72 at Oak Ridge National Laboratory
(ORNL). Computer time was provided by the Innovative and Novel
Computational Impact onTheory and Experiment (INCITE) program. This
research used resources of the Oak Ridge Leadership Computing Facility
and of the Compute and Data Environment for Science (CADES) located at
ORNL, which is supported by the Office of Science of the Department of
Energy under Contract No. DE AC05-00OR22725. R.K. acknowledges the
JSPS invitational fellowship program for short term research in Japan
at the Tokyo Institute of Technology.

\end{document}